\documentclass[twocolumn,aps,prl,showpacs]{revtex4-1}

\usepackage{hyperref}
\usepackage{amsmath}
\usepackage{color,graphicx}
\usepackage{latexsym}
\usepackage{graphics,graphicx}
\usepackage{bm,dsfont}
\usepackage{dcolumn}
\usepackage{epsfig}

\hypersetup{
    colorlinks,
    citecolor=blue,
    filecolor=blue,
    linkcolor=blue,
    urlcolor=blue
}

\begin{document}

\preprint{AIP/123-QED}

\title{Coherent manipulation of single electron spins with Landau-Zener sweeps}

\author{Marko J. Ran\v{c}i\'{c}}
\email{marko.rancic@uni-konstanz.de}
\affiliation{Department of Physics, University of Konstanz, D-78457 Konstanz, Germany}
\author{Dimitrije Stepanenko}
\email{dimitrije.stepanenko@ipb.ac.rs}
\affiliation{Institute of Physics Belgrade, University of Belgrade, Pregrevica 118, 10080 Belgrade Serbia}
\date{\today}

\begin{abstract}
We propose a novel method to manipulate the state of a single electron spin in a semiconductor quantum dot (QD). The manipulation is achieved by tunnel coupling a QD, labeled $L$, and occupied with an electron to an adjacent QD, labeled $R$, which is not occupied by an electron but having an energy linearly varying in time. We identify a parameter regime in which a complete population transfer between the spin eigenstates $|L\uparrow\rangle$ and $|L\downarrow\rangle$ is achieved without occupying the adjacent QD.
This method is convenient due to the fact that manipulation can be done electrically, without the precise knowledge of the spin resonance condition, and is robust against Zeeman level broadening caused by nuclear spins.
\end{abstract}
\pacs{03.67.Lx,73.40.Gk,81.07.Ta,71.70.Ej}
\maketitle
{\it Introduction --} Initialization, manipulation and readout of single electron spins in an efficient way are necessary for the implementation of single electron spin qubits \cite{3Loss1}.
Spin-orbit interaction and stray magnetic fields of micro-magnets provide a necessary toolkit to control the single electron spin \cite{3Flindt1,3Nowack1,3Nadj1,3Tokura1,3Pioro1,3Kawakami1}.
In Electric Dipole Spin Resonance (EDSR), microwaves drive an electron to oscillate in the spin-orbit field and/or the magnetic field gradient, producing a coherent spin rotation.

The Landau-Zener-St\"{u}ckelberg-Majorana (LZSM) model \cite{3Landau1,3Zener1,3Stueckelberg1,3Majorana1} is one of the few analytically solvable time dependent problems in quantum mechanics. It has found applications modeling nano-electro-mechanical systems \cite{3LaHaye1}, opto-mechanical systems \cite{3Heinrich1}, Bose liquids \cite{3Chen1}, molecular magnets \cite{3Wernsdorfer1}, Rydberg atoms \cite{3Rubbmark1}, superconducting qubits \cite{3Oliver1, 3Sillan1, 3Wilson1, 3Izmalkov1, 3LaHaye1} and semiconductor singlet-triplet qubits \cite{3Petta1,Petta2010,3Ribeiro1}. In the LZSM model the energy difference between two coupled states is varied linearly in time, while the coupling between the states is time independent. This results in a transition between the states with the probability determined by the coupling constant and the rate of the sweep.

Unlike the two level LZSM problem, multilevel LZSM problems are not exactly analytically solvable for a general case \cite{3Demkov1,3Carroll1,3Kayanuma1,3Pokrovsky1,3Demkov2,3Damski1,3Rangelov1}. Chirped Raman Adiabatic Passage (CHIRAP) \cite{3Oreg1,3Broers1,3Broers1,3Chelkowski1,3Shore2} and similar techniques \cite{3Gaubatz1,3Vitanov1,3Shore1,3Golter1,3Kumar1,3Bergmann1,3Bergmann2} allow for efficient transfer of populations between two uncoupled levels.
In order to utilize CHIRAP the energy of the radiatively decaying state is varied linearly in time with laser pulses having chirped frequencies.

Equivalently to CHIRAP, the goal of our scheme is to transfer the population between two
uncoupled levels $|L\uparrow \rangle$ and $|L\downarrow \rangle$ by coupling the levels of the $L$ electrostatically defined quantum dot in a time-independent manner to an adjacent electrostatically defined quantum dot, whose energy is linearly varying in time \cite{3Stehlik1}. It should be noted that, as the probability to occupy the adjacent quantum dot $R$ remains negligible in this scheme, the states in the $R$ QD can be extremely susceptible to relaxation without influencing the efficiency of our scheme. The scheme under study is also applicable to coupled donors \cite{PhysRevX.4.021044} and coupled donor-dot systems \cite{3Harvey1}. 

We discuss two possible realizations of our scheme. In the first realization the $R$ quantum dot has significantly larger Zeeman splitting than the $L$ quantum dot. Then, the scheme operates even in the case when the rate of spin-non-conserving tunneling events is significantly smaller than the rate of spin-conserving events. This regime is often present in GaAs double quantum dots. In the second realization the Zeeman splittings of the left $L$ and right $R$ quantum dots are comparable in magnitude but the rates of 
spin-conserving and spin-non-conserving tunneling events must be comparable. This regime can be reached for electrons in InAs double quantum dots and holes in GaAs double quantum dots.

\begin{figure}[t!]
	\centering
	\includegraphics[width=0.40\textwidth]{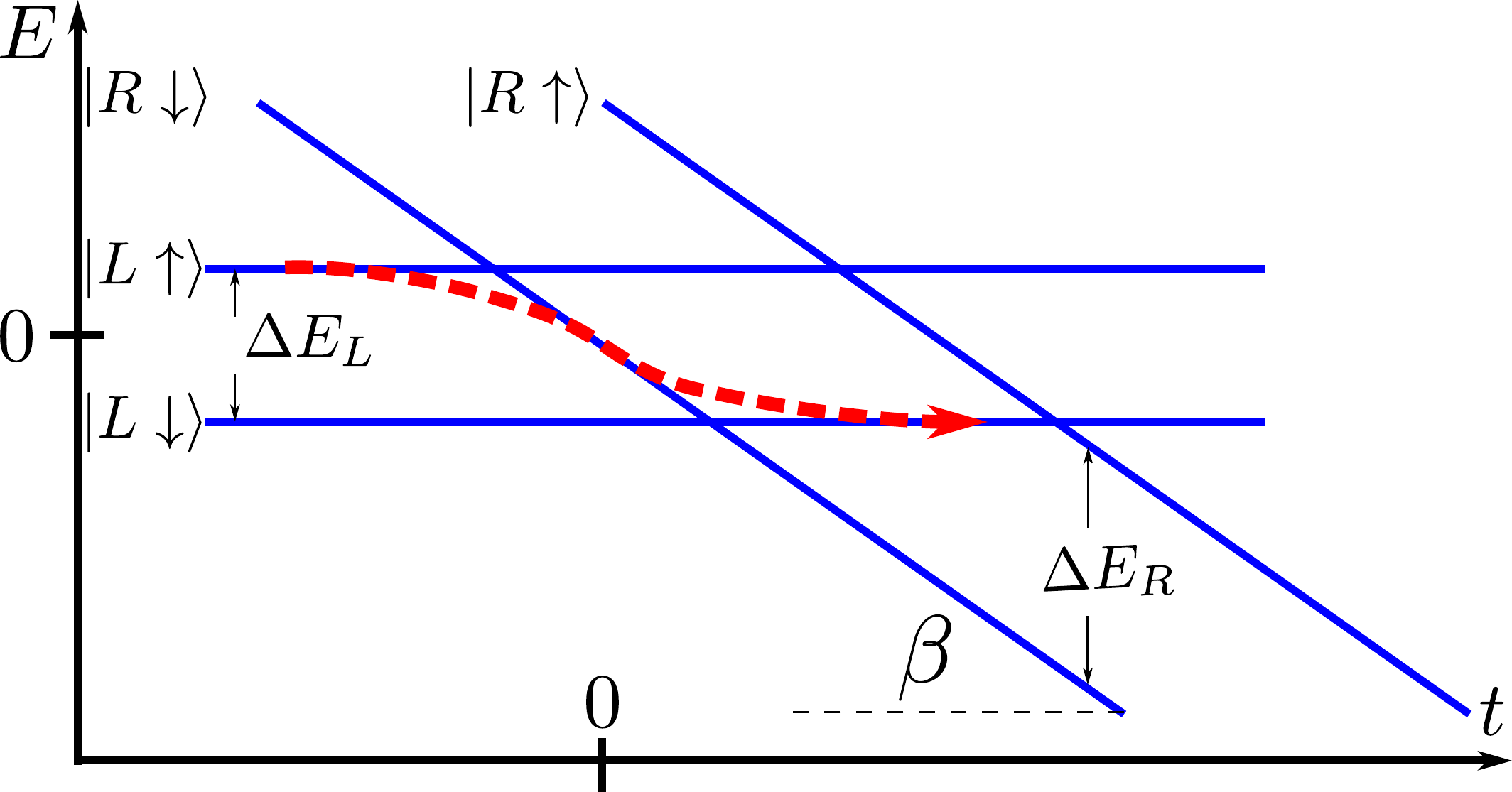}
	\caption{(Color online) The energy diagram. We initialize the electron in the $|L\uparrow\rangle$ state, with the $R$ quantum dot being higher in energy. We ramp the energies of the states in $R$ quantum dot with a Landau-Zener velocity $\beta$. In the figure $\beta<0$. The goal
	of our scheme is to find a parameter regime in which the adiabatic evolution path is followed (red dashed arrow). The Zeeman splittings of the $L$ and $R$ quantum dots are marked as $\Delta E_L$ and $\Delta E_R$ respectively.}
	\label{Setup}
\end{figure} 

{\it The Hamiltonian --} We model a situation where the electron spin is localized in the $L$ quantum dot. The energy of the $R$ quantum dot is varied linearly in time, Fig. \ref{Setup},

\begin{align}\label{eq:Ham}
H(t)&=\sum_{c}\sum_{\sigma}E_{c,\sigma}(t) |c\sigma\rangle \langle c\sigma|+\tau\sum_{\sigma}\sum_{c\neq \bar{c}}|c\sigma\rangle\langle \bar{c}\sigma| \nonumber\\
                                                                        &+\tau_\Delta\sum_{\sigma\neq \bar{\sigma}}\sum_{c\neq \bar{c}}|c\sigma\rangle\langle \bar{c}\bar{\sigma}|.
\end{align}

\noindent The sum over the charge states runs over the left and the right quantum dots, $c=L,\,R$, and the sum over spin states runs over spin-up and spin-down states $\sigma=\uparrow,\,\downarrow$. 
Furthermore, $E_{c\sigma}$ represents the energy with charge state $c$ and spin state $\sigma$. The energies of the $L$ quantum dot are time independent $E_{L\uparrow}=\Delta E_L/2$, $E_{L\downarrow}=-\Delta E_L/2$, 
where $\Delta E_L$ is the Zeeman splitting in the left quantum dot. The energies of the $R$ quantum dot are time dependent with a linear time dependence, $E_{R\uparrow}= \Delta E_R +\beta t$, and $E_{R\downarrow}=\beta t$,
where $\Delta E_R$ is the Zeeman splitting in the right quantum dot, $t$ is time and $\beta$ the Landau-Zener velocity (see Fig. \ref{Setup}).

The off-diagonal terms in the Hamiltonian are the spin-conserving tunneling amplitude $\tau$, and the spin-non-conserving tunneling amplitude $\tau_\Delta$. The spin non-conserving tunneling can appear
due to spin-orbit interaction or be induced by the stray field of the micro-magnet, which is inhomogeneous in the tunneling direction \cite{3Maisi1,3Braakman1}.
 
{\it Different Zeeman splittings --} We initialize the system in the $|L\uparrow\rangle$ state, at a negative instance of time $-T/2$. If the product of the Landau-Zener velocity $\beta$ and the total duration of the Landau-Zener sweep $T$ 
is smaller than the Zeeman splitting of the right quantum dot $\Delta E_R>\beta T$, and if the $R$ quantum dot is initially positively detuned with respect to the $L$ quantum dot, our system behaves like an effective three level system. Furthermore, if the evolution of the system is adiabatic 
($\tau^2,\,\tau_\Delta^2\gg\beta\hbar$), the system will remain in the instantaneous eigenstate of the Hamiltonian for the entire duration
of the Landau-Zener sweep $T$. Given all these assumptions, we can calculate the adiabatic eigenvectors, and therefore the time evolution of our three states probabilities

\begin{figure}[t!]
	\centering
	\includegraphics[width=0.4\textwidth]{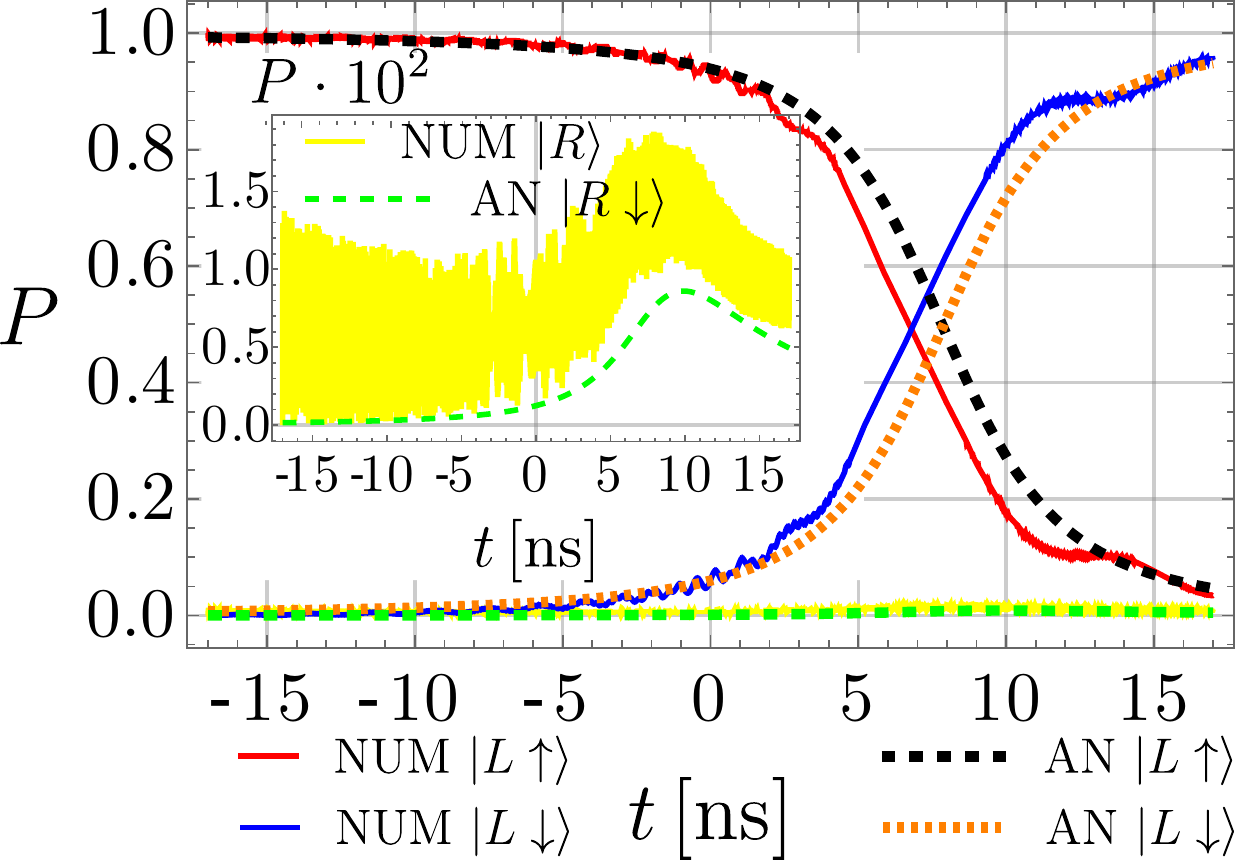}
	\caption{(Color online) The comparison between the numerically computed probabilities (obtained from evolving the state using the Hamiltonian of Eq. (\ref{eq:Ham})) ($\text{Num}$) and analytic adiabatic three level probabilities Eq. (\ref{eq:prob1}) ($\text{An}$). The parameters of the plot are the Landau-Zener velocity $\beta=5\cdot 10^3\text{ eV/s}$, the tunnel coupling $\tau=6.5\text{ $\mu$eV}$, corresponding to an interdot separation of $l=179\text{ nm}$ (for more information see Supplementary material), the spin-non-conserving tunnel coupling $\tau_\Delta=0.25\tau$, the external magnetic field Zeeman splitting in the left QD $\Delta E_L=1\text { $\mu$eV}$ and Zeeman splitting in the right quantum dot $\Delta E_R=200 \Delta E_L$. The inset represents the magnification of the occupation probabilities of the states in the $R$ quantum dot.}
	\label{Results1}
\end{figure}

\begin{gather}
P_{L\uparrow}=\tau^2_\Delta\frac{|\lambda(t)+\Delta E_L/2|^2}{{N(t)}^2},\,P_{L\downarrow}=\tau^2\frac{|\lambda(t)-\Delta E_L/2|^2}{{N(t)}^2}, \nonumber\\ 
P_{R\downarrow}=\frac{|\lambda(t)^2-\Delta E_L^2/4|^2}{{N(t)}^2},\label{eq:prob1}
\end{gather}
where $\lambda(t)$ is the appropriate adiabatic eigenvalue (see Supplementary material for the expression for $\lambda(t)$) and $N(t)$ is the normalization of the adiabatic eigenvectors.
For simplicity, we have omitted to explicitly state that $\lambda(t)$ is also a function of $\Delta E_L,\,\beta,\,\tau,\,\tau_\Delta$.
Depending on the values of $\tau$ and $\tau_\Delta$, $\lambda(t)=0$ close to $t=0$ (for $\tau=\tau_\Delta$), $\lambda(t)=0$ at $t>0$ (for $\tau>\tau_\Delta$) and 
$\lambda(t)=0$ at $t<0$ (for $\tau<\tau_\Delta$). Furthermore, the adiabatic eigenvalue takes the following values ${\lambda(t=\mp\infty)=}{\pm \Delta E_L/2}$, $-\Delta E_L/2\le\lambda(t)\le\Delta E_L/2$, for every $t$. Therefore, the maximal possible occupation probabilities are $P^{\rm max}_{L\uparrow}\sim\tau_{\Delta}^2\Delta E_L^2$, $P^{\rm  max}_{R\downarrow}\sim\Delta E_L^4$, $P^{\rm max}_{L\downarrow}\sim\tau^2\Delta E_L^2$.
If $\tau,\,\tau_\Delta\gg\Delta E_L$ 
no significant population will occupy the $R$ quantum dot, $P_{R}\approx 0$ at every instance of time (see Fig. \ref{Results1}), and a complete population transfer between the spin eigenstates $|L\uparrow\rangle$ and $|L\downarrow\rangle$ occurs.

In contrast to EDSR techniques our scheme does not require the precise knowledge of the spin resonance
condition $\Delta E_L$ and operates without microwaves. However, in order for our scheme to be successful a necessary requirement is that the quantum dots have significantly different Zeeman splittings
 $\Delta E_L\ll\Delta E_R$. For a typical double quantum dot system where the distance between the quantum dots is $\sim 200\text { nm}$ the required
 gradient would be $d B_z/dx\sim 10 \text{ T/$\mu$m}$, which is for a factor of $10$ larger than the currently maximally achieved experimental value \cite{3Obata1,3Pioro2}. A possible way to induce a large enough difference of Zeeman energies between quantum dots is to engineer the $g$-factor of one of the quantum dots $L$ to be almost zero, and engineer the $g$-factor of the $R$ QD to be significantly larger \cite{3Schroer1,3Salis1,3Jiang1,3Prechtel1}. This could be achieved by locally inducing different content of Al in the GaAs mixture \cite{3Salis1}.

{\it Equal Zeeman splittings -- }Again we initialize the system in the $|L\uparrow\rangle$ state, at a negative instance of time $-T/2$. Another way for our scheme to be successful is that the magnitude of spin-conserving and spin-non-conserving tunnelings are comparable $\tau\approx\tau_\Delta$. The requirement for our scheme to work is $\tau/\tau_{\Delta}\sim 4 l/3\Lambda_{\rm SO}\approx1$ can be fulfilled in InAs \cite{3Hensen1}. Here, $l$ is the interdot separation and $\Lambda_{\rm SO}$ is the spin-orbit length, defined by \cite{3Golovach1,3Stepanenko1}
$\Lambda_{\rm SO}=\hbar/m^*\sqrt{\cos{\phi}^2(\beta-\alpha)^2+\sin{\phi}^2(\beta+\alpha)^2},$
for a 2DEG in the $(001)$ plane. Here, $m^*$ is the effective electron mass, $\phi$ is the angle between the $[110]$ crystallographic axis and the interdot connection axis and $\beta$ and $\alpha$ are Dresselhaus and Rashba spin-orbit constants respectively.
Possible ways of controlling the spin-orbit interaction is the variation of angle between the external magnetic field
and the spin-orbit field \cite{3Nichol1}, variation of the direction in which the DQD is grown \cite{3Rancic1} (and therefore maximizing $\cos{\phi}$), isotopic control of the Indium in InGaAs, or electric field control of the Rashba constant \cite{3Nitta1,3Liang1}.

In the adiabatic limit ($\tau^2=\tau_\Delta^2\gg\beta\hbar$), the system will remain in the instantaneous eigenstate of the Hamiltonian for the entire duration
of the Landau-Zener sweep $T$. In that limit, we can calculate the adiabatic eigenvectors, and therefore the time evolution of our four states probabilities
\begin{figure}[t!]
	\centering
	\includegraphics[width=0.40\textwidth]{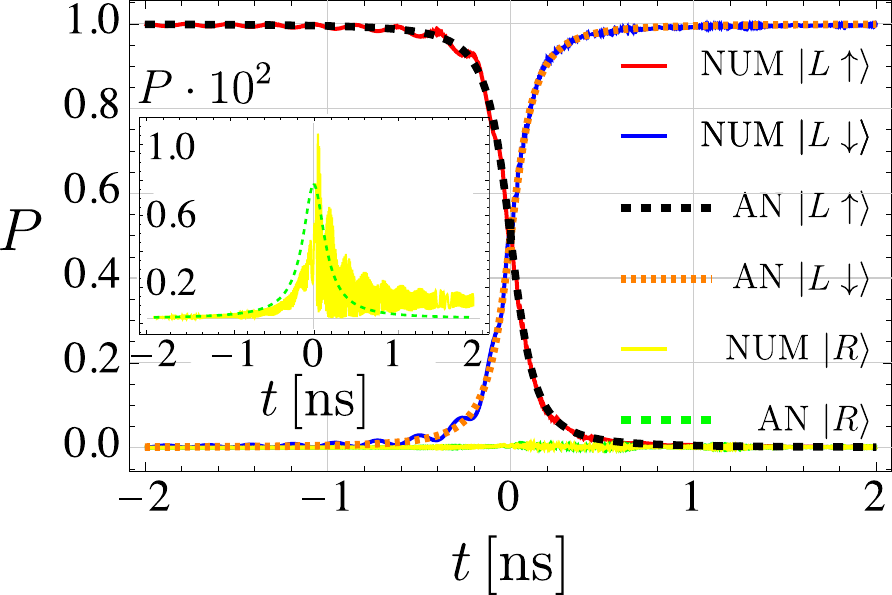}
	\caption{(Color online). The comparison between the numerically computed probabilities (obtained from evolving the state using the Hamiltonian of Eq. (\ref{eq:Ham})) ($\text{Num}$) and analytic adiabatic four level probabilities Eq. (\ref{eq:prob2}) ($\text{An}$). The inset represents the magnification of the probability to occupy the $R$ quantum dot. The parameters of the plot are the Landau-Zener velocity $\beta=4\cdot 10^6\text{ eV/s}$, the tunnel hopping $\tau=50\text{ $\mu$eV}$, corresponding to interdot distance of $l=280\text{ nm}$ for $m^*=0.023 m_e$ (for more information see Supplementary material), the Zeeman energies $\Delta E_L=\Delta E_R=17\text{ $\mu$ev}$.}
	\label{res4x4}
\end{figure}
\begin{gather}
P_{L\uparrow}=\tau^2\frac{|\Lambda(t)+\Delta E_L/2|^2}{{\tilde{N}(t)}^2},\,P_{ L\downarrow}=\tau^2\frac{|\Lambda(t)-\Delta E_L/2|^2}{{\tilde{N}(t)}^2},\nonumber\\
P_{R\downarrow}= P_{R\uparrow}=\frac{|\Lambda(t)^2-\Delta E_L^2/4|^2}{2{\tilde{N}(t)}^2},\label{eq:prob2}
\end{gather} where $\Lambda(t)$ is the corresponding adiabatic eigenvalue and $\tilde{N}(t)$ the wavefunction normalization.

  The requirement that spin-conserving and spin-non-conserving tunnel couplings are equal is due to the fact that when $\Delta E_L=\Delta E_R$ the adiabatic eigenfunctions have only a vanishing contribution of the two states of the $R$ quantum dot when $\tau\approx\tau_\Delta$ is fulfilled. In the case of $\tau\gg\tau_\Delta$, the adiabatic eigenfunctions have only a small component in the $|R\downarrow\rangle$ state when $\Delta E_L\ll\tau,\,\tau_\Delta$, and the $|R\uparrow\rangle$ state is detuned during the duration of the Landau-Zener sweep $T$.
  
  Similarly to the previous implementation of our scheme, the appropriate adiabatic eigenvalue spans between ${\Lambda(t=\mp\infty)=}{\pm \Delta E_L/2}$, $-\Delta E_L/2\le\Lambda(t)\le\Delta E_L/2$, for every $t$, with $\Lambda(t)=0$ for $t\approx 0$. 
  The maximal possible occupations of states for the case ${\Delta E_L=\Delta E_R}$ are ${P_{L\uparrow}^{\rm max.}}\sim{\tau^2\Delta E_L^2}$, ${P_{L\downarrow}^{\rm max.}\sim}{\tau^2\Delta E_L^2}$ 
  and ${P_{R\uparrow}^{\rm max.}=}{P_{R\downarrow}^{\rm max.}\sim\Delta E_L^4/2}$. Equivalently to CHIRAP, the probabilities to occupy the $|R\downarrow\rangle$ and $|R\uparrow\rangle$ states is negligible at all instances of time $P_{R}\approx0$ in the case when $\tau\gg\Delta E_L$ (see Fig. \ref{res4x4}), and a complete population transfer between the spin eigenstates $|L\uparrow\rangle$ and $|L\downarrow\rangle$ occurs.

{\it Experimental realizations -- } 
Our control scheme works optimally when the Zeeman splitting of the $L$ QD is small. Furthermore, different signs of the Landau-Zener velocity and initial detunings need to be used for different initial spin states. 
We will address the problem of initializing and measuring electron spin states when the Zeeman splitting in the $L$ QD is small in the remaining part of this subsection.

If the thermal broadening of the lead is smaller than the Zeeman splitting of the electron spin states $k_B T_e\ll \Delta E_L$, the state of the spin qubit can be initialized by tuning the chemical potential of a nearby lead close to the $|\!\downarrow\rangle$ state of the spin qubit. When lead-to-dot relaxation occurs the only possible state to which the electron can relax from the lead is the $|\!\downarrow\rangle$ state. Furthermore, single shot measurement of the electron spin state can be achieved in a similar manner \cite{3Elzerman1}, by tuning the chemical potential of the lead in such a way so that only one of the states can tunnel out of the quantum dot to the lead.

As our scheme operates optimally in low magnetic fields $k_B T_e>\Delta E_L$, the initialization and readout, validating the efficiency of our scheme, must be done in an alternative way, via the $R$ QD. The chemical potential of the lead coupled to the $R$ QD can be tuned between the spin states of the $R$ QD. After the successful initialization the $|R\downarrow\rangle$ state, the spin is shuttled to the $|L\downarrow\rangle$ state, followed by a manipulation of the spin according to our scheme. 

After the manipulation stage the modification in the current of a quantum point contact (QPC) near to $R$ is monitored. If the current of the QPC is unchanged, this means that the manipulation stage did not produce any leakage to the $R$ quantum dot and that the spin measurement stage can follow. In the spin measurement stage states $|L\downarrow\rangle$ and $|R\downarrow\rangle$ are aligned in energy one more time. If the electron spin was in the $|L\downarrow\rangle$ state a tunneling event occurs and a near by QPC modifies its current accordingly \cite{3Barrett1,3Shin1}. On the other hand if the electron spin was in the $|L\uparrow\rangle$ state the current of the QPC would remain unchanged.

In the case of $\Delta E_L= \Delta E_R$ (and therefore $\tau \approx\tau_\Delta$) and when $\Delta E_L<k_BT_e$ the initialization could still be achieved by waiting a sufficiently long time for the electron spin to relax to the thermal equilibrium state. However, spin readout would need to be done with alternative methods, because both spin eigenstates are energetically allowed to tunnel to the $R$ QD when $|L\downarrow\rangle$ and $|R\downarrow\rangle$ are aligned in energy. This is why we consider the case $\Delta E_L\ll \Delta E_R$ to be more likely to implement in future experiments, and only consider the influence of nuclear spin noise for this realization.

\begin{figure}[t!]
	\centering
	\includegraphics[width=0.46\textwidth]{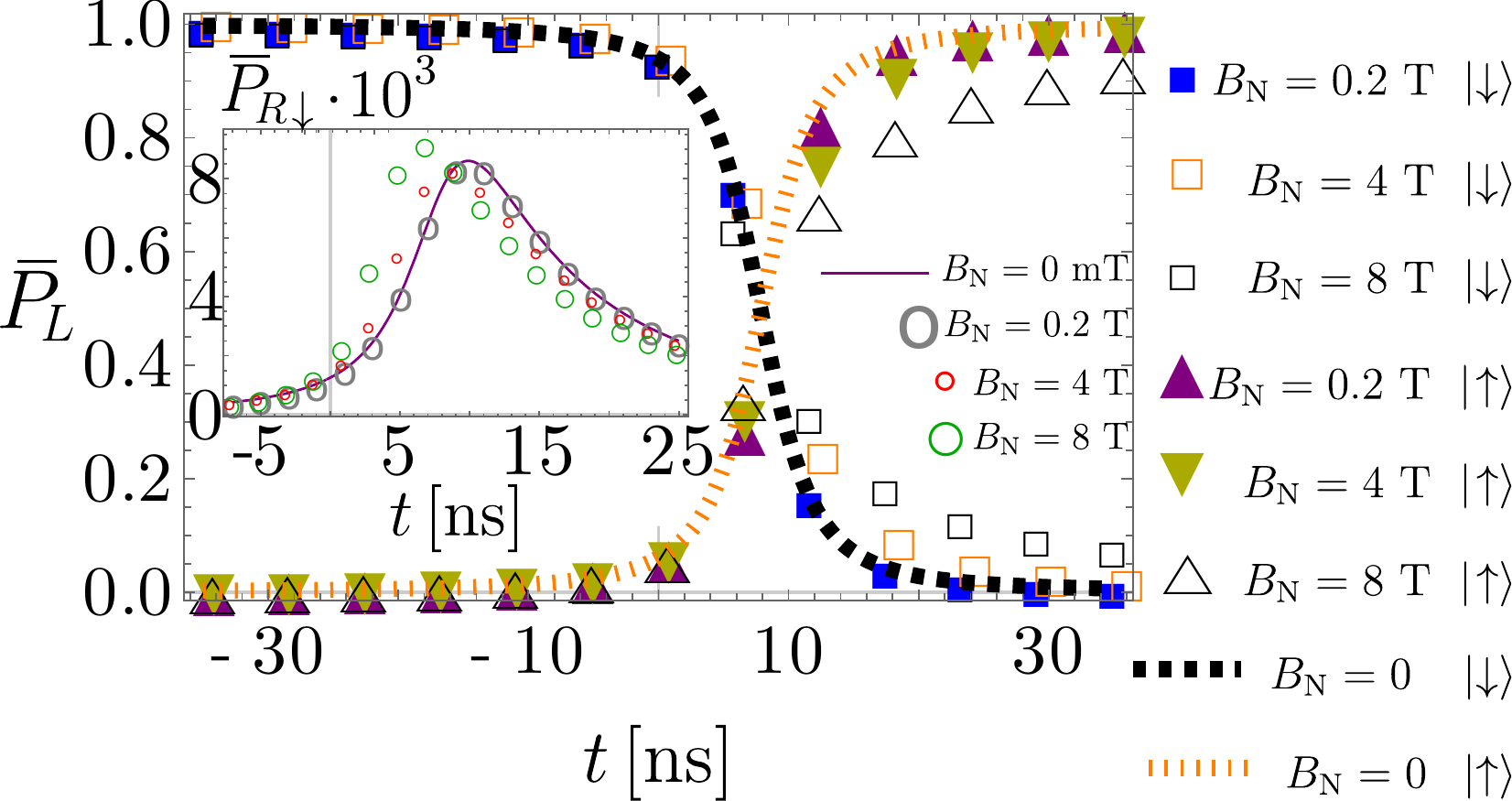}
	\caption{(Color online) Spin manipulation in the presence of nuclear spins. The parameters of the plot are the Landau-Zener velocity $\beta=5\cdot 10^3\text{ eV/s}$, the tunnel coupling $\tau=6.5\text{ $\mu$eV}$, corresponding to an interdot separation of $l=179\text{ nm}$ (for more information see Supplementary material), the spin-non-conserving tunnel coupling $\tau_\Delta=0.25\tau$, the Zeeman energy in the left quantum dot $\Delta E_L=1\text{ $\mu$eV}$, the standard deviation in the right quantum dot $\chi=0$, the $g$-factor in the left quantum dot $g_L=1.2\cdot 10^{-3}$. The inset represents occupation of the states in the $R$ quantum dot.}
	\label{NS1}
\end{figure}

{\it Errors due to nuclear spins -- }
We model the influence of nuclear spins as a distribution of the magnetic field in the $L$ and $R$ quantum dot, centered around the external magnetic field in the left and the right dot $\Delta E_L$, $\Delta E_R$, with standard deviations $\sigma=g_L\mu_BB_N$, $\chi=g_R\mu_BB_N$, where $g_{L(R)}$ is the electron $g$-factor in the left (right) quantum dot, $\mu_B$ is the Bohr magneton and $B_N$ is the root-mean-square of the distribution of the nuclear magnetic field \cite{PhysRevB.65.205309}.
The influence of nuclear spins on our manipulation scheme can be estimated by averaging the probabilities of all relevant states over a distribution of nuclear spins

\begin{figure}
		\centering
		\includegraphics[width=0.46\textwidth]{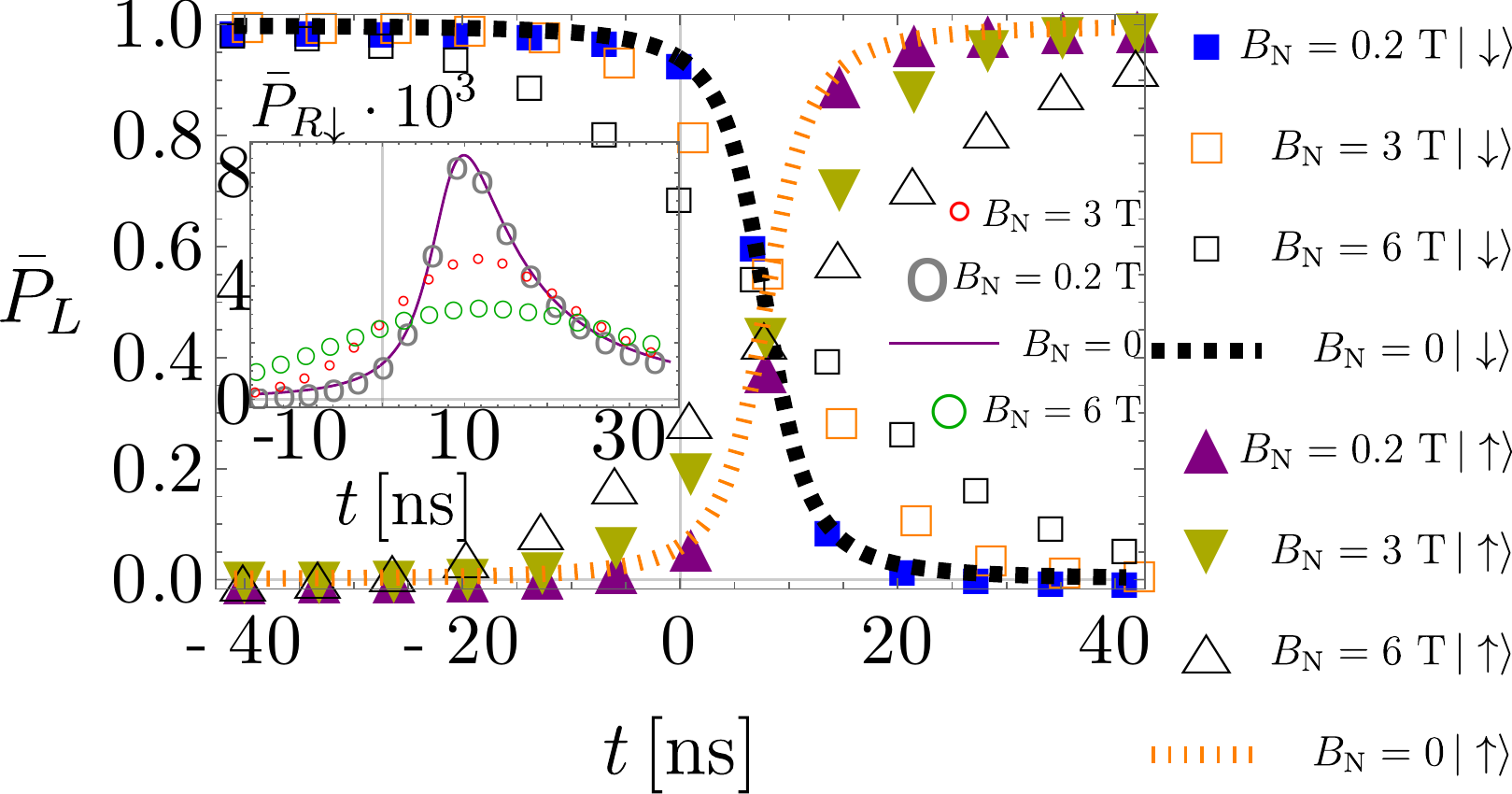}
		\caption{(Color online) Spin manipulation in the presence of nuclear spins. The parameters of the plot are the Landau-Zener velocity $\beta=5\cdot 10^3\text{ eV/s}$, the tunnel coupling $\tau=6.5\text{ $\mu$eV}$, corresponding to an interdot separation of $l=179\text{ nm}$ (for more information see Supplementary material), the spin-non-conserving tunnel coupling $\tau_\Delta=0.25\tau$, the Zeeman splitting in the left quantum dot $\Delta E_L=1\text{ $\mu$eV}$, the $g$-factor in the left quantum dot $g_L=1.2\cdot 10^{-3}$, the $g$-factor in the right quantum dot $g_R=200g_L$. The inset represents occupation of the states in the $R$ quantum dot.}
		\label{NS2}
\end{figure}

\begin{equation}
\bar{P}_{c\sigma}=\iint\limits^{\infty}_{-\infty}\frac{P_{c\sigma}}{2\pi \chi\sigma} e^{-\frac{\left(\Delta E-\Delta E_L\right)^2+\left(\tilde{\beta} \tilde{t}-\beta t\right)^2}{4\sigma^2\chi^2}}d(\Delta E) d \left( \tilde{\beta} \tilde{t}\right)
\end{equation}
where $c=L,\,R$, $\sigma=\uparrow,\,\downarrow$, with the exclusion of the detuned $|R\uparrow\rangle$ state.

In Fig. \ref{NS1} we show how the nuclear spins influence our control scheme in the case of no uncertainty of the magnetic field in the right quantum dot, $\chi=0$. If the random nuclear field is parallel with the external magnetic field this gives rise to more leakage into the $|R\downarrow\rangle$ state. However, if the random nuclear field is anti-parallel with the external magnetic field this gives rise to less leakage into the $|R\downarrow\rangle$ state, and this two effects (less and more leakage to $|R\rangle$) cancel first order in $\Delta E_L$.

In Fig. \ref{NS2} we present the behavior of our control scheme under an influence of random nuclear spins in both quantum dots. Other then the already mentioned mechanism of additional leakage, the uncertainties in the nuclear field in the right quantum dot (and therefore the position of the level $|R\downarrow\rangle$) lead to reduced maximal probability to occupy the $|R\downarrow\rangle$ state (Fig. \ref{NS2}, inset, dark gray versus green circles). In contrast to EDSR, we are able to achieve a full transfer of population between the spin eigenstates, even when the uncertainty in the energy difference between spin eigenstates is large (Fig. \ref{NS2} black empty squares and triangles).

An effective nuclear magnetic field of unknown intensity in the $z$ direction is going to change the instance of time in which the energy of the state $|R\downarrow\rangle$ is located between the energies of the states $|L\uparrow\rangle$ and $|L\downarrow\rangle$. For a nuclear magnetic field parallel with the external field the energy of the state $|R\downarrow\rangle$ is located between the energy of the states $|L\uparrow\rangle$ and $|L\downarrow\rangle$ at a time $t<0$. In contrast to that, for a nuclear magnetic field anti-parallel with the external field the energy of the state $|R\downarrow\rangle$ is located between the energies of the states $|L\uparrow\rangle$ and $|L\downarrow\rangle$ at a time $t>0$. A process like this is described with a Gaussian distribution, centered around $\beta t$ with a standard deviation $\chi=g_R\mu_BB_N$ where, $g_R$ is the $g$-factor in the right quantum dot, $g_R\gg g_L$. This leads to a reduced maximal value of the occupation of the $|R\downarrow\rangle$ state, without changing the averaged occupation of the $|R\downarrow\rangle$ per unit time 
$\bar{\bar P}_{R\downarrow}(T)=\int_{-T/2}^{T/2} \bar{P}_{R\downarrow}(t) dt/T=\text{const.}$ for a large enough $T$. Since the nuclear spins do not affect the final probabilities, our scheme can be operated in the presence of nuclear spin induced decoherence, as long as the total sweep time (in our case $\sim 80\text{ ns}$) is shorter than the characteristic time of nuclear spin evolution ($1\text{ $\mu$s}$) \cite{PhysRevB.65.205309}.
In should be noted that quasi-static detuning noise yields the same effect like having an uncertain nuclear spin distribution in the $R$ quantum dot, and therefore we do not address this issue separately in this manuscript.

{\it Conclusions and final remarks -- } To conclude, we have proposed a novel method to manipulate a single electron spin by using Landau-Zener sweeps. Our control method is robust against the uncertainties of the nuclear field and static charge noise, operates without microwaves and without the precise knowledge of the spin resonance condition. We thank Marko Milivojevi\'{c}, Guido Burkard, Maximilian Russ, Alexander Pearce and Ferdinand Kuemmeth for fruitful discussions. This work is funded from MPNTR grant OI171032, DAAD grant 451-03-01858201309-3, and European Union within the S$^3$nano initial training network.
\bibliographystyle{apsrev}
\bibliography{References3}

\end{document}